\documentstyle[12pt]{article}
\def\+{{+\!\!\!+}}
\def\pp{\mbox{\tiny${}_{\stackrel\+ =}$}}
\def\mm{\mbox{\tiny${}_{\stackrel= \+}$}}

\def\cA{{\cal A}}

\def\cM{{\cal M}}

\def\d{\partial}

\def\a{\alpha}
\def\th{\theta}
\def\g{\gamma}
\def\G{\Gamma}

\def\N{\nabla}

\def\P{\Psi}

\def\h{\chi}

\def\r{\rho}
\def\l{\lambda}

\def\L{\Lambda}

\def\s{\sigma}

\def\e{\epsilon}
\def\ve{\varepsilon}

\def\is{\equiv}

\def\half{{\textstyle{1 \over 2}}}
\def\ihalf{{\textstyle{i \over 2}}}
\def\fou{{\textstyle{1 \over 4}}}

\def\bop#1{\setbox0=\hbox{$#1M$}\mkern1.5mu
        \vbox{\hrule height0pt depth.04\ht0
        \hbox{\vrule width.04\ht0 height.9\ht0 \kern.9\ht0
        \vrule width.04\ht0}\hrule height.04\ht0}\mkern1.5mu}

\newcommand\hv[1]{(\ref{#1})}

\begin{document}

\newcommand{\inv}[1]{{#1}^{-1}} 

\renewcommand{\theequation}{\thesection.\arabic{equation}}
\newcommand{\beq}{\begin{equation}}
\newcommand{\eeq}[1]{\label{#1}\end{equation}}
\newcommand{\ber}{\begin{eqnarray}}
\newcommand{\eer}[1]{\label{#1}\end{eqnarray}}
\begin{center}
                        \hfill    USITP-98-13\\
                        \hfill    July 1998\\

\vskip .4in \noindent

\vskip .1in

{\large \bf A Picture of $D$-branes at Strong Coupling II.\\
Spinning Partons}
\vskip .2in

{\bf Henrik Gustafsson}$^{a,}$\footnote{e-mail: henrik@physto.se} and
{\bf Ulf Lindstr\"om}$^{a,b,}$\footnote{e-mail: ul@physto.se}
\bigskip \\
$\mbox{}^{a)}$ {\em  Institute of Theoretical Physics,
University of Stockholm \\
Box 6730,
S-113 85, Stockholm, SWEDEN}\\
\bigskip

$\mbox{}^{b)}$ {\em Institute of Physics, University of Oslo\\
Box 1048, N-0316 Blindern, Oslo, NORWAY\\}
\vskip .15in

\vskip .1in
\end{center}
\vskip .4in
\begin{center} {\bf ABSTRACT } \end{center}
\begin{quotation}\noindent
We study the Born-Infeld  $D$-brane action in the limit $g_s\to \infty$.
The
resulting actions is presented in an arbitrary background and shown to
describe a foliation of the world-volume by strings. Using a recently
developed ``degenerate'' supergravity the parton picture is shown to
be
applicable also to supersymmetric $D$-branes.
\end{quotation}
\vfill
\thispagestyle{empty}\eject
\setcounter{page}{1}

\section{Introduction}
In a recent paper a picture of $D$-branes at strong coupling was
presented \cite{lind1}. There it was shown that $D$-branes can be
viewed as bound states of string-partons in a certain limit. More
precisely, the limit when the string coupling $g_s$ goes to
infinity with $\a^\prime$ kept fixed corresponds to the limit
when the $Dp$-brane tension $T_p$ goes to zero. In this limit the
Born-Infeld action for the $Dp$-brane is replaced by
\beq
S={\textstyle{1\over 4}}\int d^{p+1}\xi V^iW^j(\g_{ij}+F_{ij}),
\eeq{VWac}
where $\g_{ij}+F_{ij}$ is the induced metric and world-volume field
strength plus antisymmetric tensor, and
$V^i$ and
$W^i$ are world-volume vector densities replacing the auxiliary
metric. The solutions to the field equations that follow from
(\ref{VWac}) can be shown to describe strings in a particular gauge
where two of the world-volume coordinates act as string coordinates
and the rest just become labels. It is this ``foliation'' we have in
mind when we say that $D$-branes at strong coupling are described by
string-partons.

Now, in \cite{lind1} only the bosonic part of the $D$-branes was
considered. In general one would like to carry out a similar
analysis for the full space-time supersymmetric theory. This would
presumably lead to super-strings as partons. Equivalently these
could be viewed as spinning (world-sheet supersymmetric) strings.
Note that although there is no reason to expect that the
supersymmetric $D$-branes have an equivalent spinning version
this is nevertheless true for the string-partons. It will
therefore serve as a consistency check of our picture if we can
find a ``spinning'' version of (\ref{VWac}). We use quotation marks
since we are not talking about a standard $(p+1)$-dimensional
supergravity version, but rather a ``degenerate'' type of
supergravity\footnote{Degenerate in that the moving frames are
non-invertible.}
which leads to the partons being spinning strings. The problem very much
resembles that encountered in finding the spinning version of the
tensionless
(fundamental) string
\cite{lst1}. In a certain gauge, the tensionless string may be viewed as
a
collection of massless particles moving subject to a constraint.
World-sheet
supersymmetrization is effected by coupling the string fields to a
``degenerate'' $2D$-supergravity and leads to the spinning
tensionless string being described by a collection of spinning
particles \cite{lr}. This procedure
is best described in a  superspace formulation, and in
this letter we present a similar approach to the $D$-brane limit
(\ref{VWac}).

The plan of the
paper is as follows: Section 2 contains our main results, including the 
action for a spinning version of $D$-branes at strong coupling. In
Section 3 we
recapitulate the derivation of the $T_p\to 0$ limit of \cite{lind1} and
extend
it to include a nontrivial background. In  Section 4 we give a short
presentation of the new supergravity and apply it to construct the
spinning version of the $D$-branes. Section 5 contains our conclusions.

\section{Results}
Some time ago one of the authors showed how to write a first order
version of the Born-Infeld action for a bosonic {\em p}-brane,
\beq
S^2_{BI}(T_{p})=T_{p}\int d^{p+1}\xi \sqrt{-\det(
       \g_{ij}+F_{ij})},
\eeq{binf}
where $\g _{ij}\equiv \d _iX^\mu \d_jX^\nu G_{\mu\nu}(X)$ is the
 metric on the world-sheet induced from a background metric $G_{\mu\nu}$
and
$F_{ij}$ is a world-volume field strength, which in this article we will
take
to be
\beq
F_{ij}\equiv \d_{[i}A_{j]}+B_{ij} .
\eeq{fdef}
Here $B_{ij}\equiv\d_iX^\mu\d_jX^\nu B_{\mu\nu}$ is the pull-back of the
background Kalb-Ramond field. The
first order action is \cite{ul}:
\beq
S^1_{BI}(T)=\half T\int
d^{p+1}\xi \sqrt{-s}\left( s^{ij}(\g _{ij}+F_{ij})-(p-1)\right),
\eeq{sbinf}
where $s^{ij}$ is a general second rank world-volume
tensor, (no symmetry as\-sumed)\footnote{This form of the action has
recently
been used to discuss the geometry of $D$-branes
\cite{Hull1}, to investigate the rigid symmetries of
$D$-branes \cite{Brandt} and as a starting point for a Hamiltonian
discussion of $D$-strings \cite{Lee}.}. 
We take \hv{sbinf} to represent the $D$ brane action,
thus disregarding the overall dilaton factor $e^{-\phi}$ in the
Lagrangian, since it will play no role for our considerations. If one
wishes, it is
readily reinserted in all our formulae.

In \cite{lind1} (\ref{sbinf}) was derived from (\ref{binf}) via a
Hamiltonian
formulation. This process also allowed the limit $T\to 0$ to be taken
and thus
the formulation (\ref{VWac}) was found. An equivalent formulation of the
tensionless
$D$-brane action was also given:
\beq
S^1_{BI}(0)={\textstyle{1\over 4}}\int d^{p+1}\xi
(\eta^{AB}e^i_Ae^j_B-\ve^{AB}e^i_{A}e^j_{B})(\g_{ij}+F_{ij}),
\eeq{Eac}
where $A,B=0,1$, $e_A\is e^i_A\d_i$ are ``degenerate'', (for $p>1$),
zwei-beins
corresponding to a $2D$ Lorentzian ``tangent space''.
In that tangent space the Minkowski metric is
$\eta^{AB}$ and $\ve^{AB}$ is the epsilon symbol.
It is this form that will serve as our starting point for
supersymmetrization.
The form of the actions \hv{Eac} and (\ref{VWac}) clearly show the
nature
of
the strong coupling limit of $D$-branes that we are considering: {\em In
this
limit
the
$D$-brane dynamics is governed by actions that involve a ``degenerate''
metric 
of rank $2$ ,
($\propto e^i_Ae^j_B\eta^{AB}$).}

Supersymmetrization of \hv{Eac} entails promoting the target space
coordinates
$X^\mu(\xi)$ to superfields $X^\mu(\xi,\th)$ and coupling them to a
supergravity.
The supergravity we use is a recently developed ``degenerate''
supergravity
\cite{gulin} whose basic superfield-densities are $E^\cM_\cA$ with $\cM
\in
\{+,-,i\},
\, i=0,\ldots,p$ and $\cA \in \{+,-,\+ ,=\}$. The bosonic densities in
\hv{Eac} are
given by
\beq
e\pp^i \is E\pp^i |
\eeq{Ecom}
where $|$ denotes ``the
$\th$-independent part of''. The spinning version of \hv{Eac} reads
\beq
{\bf S}_{BI}=-\fou\int d^{p+1}\xi d^2\th \N_+X^\mu\N_-X^\nu {\cal
E}_{\mu\nu}(X),
\eeq{Spinac}
where $\N_\pm \is E_\pm^\cM\d_\cM+\omega_\pm M$, with $M$ the $2D$
Lorentz
generator, and $ {\cal E}_{\mu\nu}\is G_{\mu\nu }+B_{\mu\nu}$.  Below
we shall
once again restrict to a flat background. Note that there is no $A_i$
field in
\hv{Spinac}. In fact, a feature of the supergravity constraints is that
the
$A_i$ field equations that follow from \hv{Eac} are automatically
satisfied. 
The action \hv{Spinac} is thus the  supersymmetization of \hv{Eac} with
the $A_i$
field  integrated out. We relegate the details of this model to Section
4 below
and end this section by presenting the component version of \hv{Spinac}:
\ber
{\bf S}_{BI}&=&\fou\int
d^{p+1}\xi\left\{\d_\+X\cdot\d_=X+2\h_=^{\ \ +}\d_\+X\cdot\Psi_+
+2\h_\+^{\ \ -}\d_=X\cdot\Psi_-\right.\cr
&&\left.+i\Psi_+\cdot\d_=\Psi_+-i\Psi_-\cdot\d_\+\Psi_-
-2\left(\Psi_+\cdot\Psi_-\right)\left(\h_=^{\ \ +}\h_\+^{\ \
-}\right)\right.\cr
&&\left. +{\cal F}\cdot{\cal F}-A_i\d_j(e_{[\+}^ie_{=]}^j)\right\}\cr
&=&\fou\int d^{p+1}\xi
\left\{\eta^{AB}e^i_Ae^j_B\d_iX\cdot\d_jX
+2\bar{\h}_A\g^i\g^A\Psi\cdot\d_iX\right.\cr
&&\left.+i\bar{\Psi}\cdot\g^i\d_i\Psi-\half\left({\bar\Psi}\cdot
\Psi\right)\left({\bar\h}_A\g^B\g^A\h_B\right)\right.\cr
&&\left. +{\cal F}^{\alpha\beta}\cdot {\cal
F}_{\alpha\beta}-\ve^{AB}A_i\d_j(e_{A}^ie_{B}^j)\right\}.
\eer{Spicom}
Here $\h $ is the supergravity spinor (density)
field, $\Psi$ and ${\cal F}$ are the spinor and auxiliary partners of
$X$
and $\cdot$ denotes target space contraction 
with
the flat metric. Note that we have reintroduced $A_i$ as a Lagrange
multiplier
to be able to work with unconstrained $e_A^i$'s. To facilitate
comparison to the
spinning string we have also included the covariant version, introducing
the
$2D$ 
$\g$-matrices and
denoting $2D$ spinor indices by $\alpha ,\beta ,...$
\footnote{Here ${\cal F}_{\alpha\beta} =\ve_{\alpha\beta}{\cal F}$.}.
The only
field that is not a density is $X$. For the special case of $p=1$ we do
indeed
recover the spinnig string, albeit with redefined fields to take care of
their
density characters.

\section{Derivation of the bosonic model}
In this section we briefly recapitulate the derivation of \hv{sbinf}
from
\hv{binf}
for arbitrary $p$, extending it to include a non-trivial
background\footnote{ In
\cite{lind1} the derivation is given for
$G_{\mu\nu}=\eta_{\mu\nu}$ and $B_{\mu\nu}=0$.}.  We follow the
procedure
originally presented in \cite{robert}, and described in this context,
e.g., in
\cite{tless}, i.e., we derive the momenta, the constraints and then the
Hamiltonian. Integrating out the momenta
from the phase space Lagrangian we then obtain a configuration space
action
with
the Lagrange multipliers for the constraints among the variables. We
finally
identify those multipliers with geometric objects on the 
world volume\footnote{The Hamiltonian analysis of \cite{lind1} was
used in covariant quantization of $D$-branes \cite{Kallosh} and 
the phase space Lagrangian was used in studying solitonic 
solutions of branes within branes \cite{Gauntlett}.}.

The generalized momenta that follow from \hv{binf} are (dropping the $p$
index
on $T$)
\ber
\Pi_{\mu} &=& \half T\sqrt{-\det(\sigma_{kl})}
\partial_{i}X^{\sigma}\left[\sigma^{(i0)}
G_{\sigma\mu}+\sigma^{[0i]}B_{\sigma\mu}\right]\cr
P^{a} &=& \half T\sqrt{-\det(\sigma_{kl})}
  \sigma^{[a0]},\cr
P^{0} &=& 0,
\eer{mom}
where $\sigma_{ij} \equiv \g_{ij}+F_{ij}$ and $\s^{ij}$ is its inverse.
The primary constraints are

\ber
 \tilde\Pi_{\mu}\partial_{a} X^{\mu} + P^{b}F_{ab} &=& 0,\cr
 P^{0} &=& 0,\cr
 \tilde\Pi_{\mu}G^{\mu\nu}\tilde\Pi_{\nu} + P^{a}\gamma_{ab}P^{b}
+ T^2\det\left(\s_{ab}\right) &=& 0,
\eer{con1}
where $a,b,\ldots =1,\ldots ,p$ are transversal indices and
$\tilde\Pi_\mu\equiv \Pi_\mu+P^aB_{a\mu}$. With the substitution
$\Pi\to\tilde\Pi$ and the full background $F_{ij}$ from \hv{fdef}, the
constraints
\hv{con1} are formally identical to those in a trivial background.  The
corresponding Hamiltonian is
\ber
 {\cal H} &=&
 P^{i}\partial_{i} A_{0} + \chi P^{0}
+\rho^{a}\left(\tilde\Pi_{\mu}\partial_{a}
  X^{\mu} + P^{b}F_{ab}\right) \cr
  && +\lambda \left( \tilde\Pi_{\mu}G^{\mu\nu}\tilde\Pi_{\nu}
  +P^{a}\gamma_{ab}P^{b} + T^2\det\left(\s_{ab}\right)
   \right),
\eer{h1}
where $\chi ,\r^a$ and $\l $ are Lagrange multiplers for the
constraints.
Including a secondary "Gauss' law"
type constraint,
the final Hamiltonian is just the sum of the
constraints, in agreement with the diffeomorphism invariance of the
original
Lagrangian, and
the phase space action may be written
\ber
S_{PS} &=& \int d^{p+1}\xi \left[\tilde\Pi_{\mu}\partial_{0}X^{\mu}
\right.
  + \left. P^{a}F_{0a} - \chi
P^{0}\right.\cr
  &&- \lambda\left(\tilde\Pi_{\mu}G^{\mu\nu}\tilde\Pi_{\nu} +
P^{a}\gamma_{ab}P^{b} +
     T^2\det\left(\s_{ab}\right)\right)\cr
 &&- \left.\rho^{a}\left(\tilde\Pi_{\mu}\partial_{a}X^{\mu} +
P^{b}F_{ba}
   \right)\right].
\eer{sps}
This is identical to the $B_{\mu\nu}=0$ phase space action presented in
\cite{lind1}, (with $\Pi \to\tilde\Pi$)\footnote{This change of
coordinates in
phase space has a trivial Jacobian.}. The analysis proceeds as in that
paper
from
here on: integrating out
$\tilde\Pi$ and
$P$, linearizing the resulting action and finally identifying the
Lagrange
multipliers with
$s_{ij}$ we obtain \hv{sbinf}.

The derivation of the $T\to 0$ limit also follows directly as in
\cite{lind1}:
the result is \hv{VWac} or, equivalently, \hv{Eac} (now including the
full
background).
\bigskip

We would also like to recapitulate how the string-parton picture arises.
To
contribute something new, we will again assume a non-trivial background.
The
equations of motion that follow from variation of (\ref{VWac}) with
respect to
$W^i$,
$V^i$,
$A_i$ and
$X^\mu$ are
\ber
 V^{i}\left(\gamma_{ik} + F_{ik}\right) &=& 0,\cr
 \left(\gamma_{ik} + F_{ik}\right)W^{k} &=& 0,\cr
 \partial_{i}\left(V^{[i}W^{k]}\right) &=& 0,\cr
\partial_{i}\left[\left(V^{(i}W^{k)}G_{\mu\nu}+
V^{[i}W^{k]}B_{\mu\nu}\right)\partial_{k}X^{\nu}\right]&&\cr
-V^iW^j\partial_iX^\rho\partial_jX^\nu(G_{\rho\nu}+B_{\rho\nu}),_\mu&=&
0,
\eer{eom}
The first two equations of motion can be reduced to
\ber
 V^{i}V^{k}\gamma_{ik} &=& 0,\cr
 W^{i}W^{k}\gamma_{ik} &=& 0,
\eer{sil2}
which say that $V$ and $W$ are null-like vector fields in the induced
metric.
The third equation can be written as $\left[V,W\right]^{i} =
(\partial\cdot W) V^{i} - (\partial\cdot V) W^{i}$, which, after
choosing
the gauge $\partial\cdot V = \partial\cdot W = 0$ says that $V$ and $W$
commute and define good coordinates. The coordinates thus defined
coordinatize two dimensional submanifolds of the world volume. They
will be the world sheets of the constituent strings. On each 
world-sheet we have the differential operators
\ber
 \partial_{\+} &=& V^{i}\partial_{i},\cr
 \partial_{=} &=& W^{i}\partial_{i}.
\eer{sil3}
Using these and the gauge choice we see that the equations of motion
reduce to
\ber
 \gamma_{\+\+} &=& 0,\cr
 \gamma_{==} &=& 0,\cr
 \partial_{\+}\partial_{=}X^{\mu}
+\Gamma_{\nu\rho}^{~~\mu}\partial_{\+}X^\rho\partial_{=}X^{\nu}&=& 0,
\eer{sil4}
where $\Gamma_{\nu\rho}^{~~\mu}$ is the torsionful connection expressed
in
terms of
$G_{\mu\nu}$ and $B_{\mu\nu}$. The equations \hv{sil4} are exactly the
conformal
gauge equations of motion for a string in a non-trivial background. The
additional coordinate dependence  of $X^{\mu}$ now becomes a
label distinguishing different string world-sheets.

For the special case when $V^i$ and $W^i$ are parallel, an analogous
analysis shows that the world volume splits into a collection of
massless particles.

\section{Derivation of the supersymmetric model}

In this Section we briefly present the ``degenerate'' $(p+1)$-dimensional 
supergravity. The structure is reminiscent of, but still distinctly
different
from the usual $2D$ superspace supergravity, as described in, e.g.,
\cite{jim},
\cite{rvnz}. Details can be found in
\cite{gulin}. We also discuss the  supersymmetrisation of
\hv{Eac} leading to \hv{Spinac} in some detail.

The defining relations for our supergravity are
\ber
\N_\pm &=& E_\pm^+\d_++E_\pm^-\d_-+E_\pm^i\d_i+\omega_\pm M,\cr
\N\pp &=& e\pp ^i\d_i+\h\pp^+\d_++\h\pp^-\d_-+\omega\pp M,
\eer{nabla}
along with the constraints
\ber
&&\{\N_+,\N_-\}+\G_{(+}\N_{-)}=RM,\cr
&&\N^2_\pm+\G_\pm\N_\pm=\pm i\N_{\pp}.
\eer{scons}
Here the additional ``connections'' are given by
\beq
\G_\pm\is \d_i E^i_\pm+\d_+ E^+_\pm+\d_-
E^-_\pm\pm\half\omega_\pm\is1\cdot{\stackrel \leftarrow\N}_\pm.
\eeq{gamma}
All fields are superfields and depend on the superspace coordinates $\xi
^i,\th^+$ and $\th^-$. The partial derivatives are defined with respect
to those
coordinates. A novel feature is that the
$\th$'s transform as (weight
-$\fou$) densities under $\xi$ diffeomorphisms. Diffeomorphisms,
($\s^i$), 
supersymmetry, ($\e^\pm$) and Lorentz, ($\L$),
transformations are coded into the superfield
$K$ defined by
\beq
K\is \s ^i\d_i+\e^+\d_++\e^-\d_-+\Lambda M,
\eeq{Kdef}
and the transformations of the derivatives in \hv{nabla} are given by
\ber
\delta \N_\pm &=& [\N_\pm ,K]-\half (1\cdot{\stackrel \leftarrow
K})\N_\pm \, , \cr
\delta \N_{\pp} &=& [\N_{\pp} ,K]- (1\cdot{\stackrel
\leftarrow K})\N_{\pp}  \, ,
\eer{natra}
where
\beq
1\cdot{\stackrel
\leftarrow K}\is\d_i\s^i-\d_+\e^+-\d_-\e^-.
\eeq{Kback}
These are the appropriate transformations for densities of weights
$\fou$ and
$\half$ respectively. From these relations the transformations of the
components may be derived. 
To display the physical content of the theory it is
convenient to work in a Wess-Zumino (WZ) gauge which we define as
follows:
\beq
\N_\pm|=\d_\pm, \qquad \left[\N_\pm
,\N_\mp\right]|+\G_{[\pm}\N_{\mp]}|=0.
\eeq{wz}

From the constraints and the Bianchi identities it follows that only
certain
component fields are independent.
We define components by projection and use the
same notation for the supergravity superfields and their lowest
components:
\ber
e\pp^i &\is& e\pp^i |, \qquad \h\pp^\pm\is\h\pp^\pm| ,
\qquad
\omega\pp\is\omega\pp|,\cr
R&\is& R|, \qquad \rho_\pm \is \N_\pm R|.
\eer{sgcom}
We shall also need the components of a scalar superfield $X$ (in
WZ-gauge),
\beq
X^\mu\is  X^\mu |, \qquad \Psi_\pm^\mu\is\d_\pm X^\mu |,\qquad {\cal
F}^\mu\is\d_+\d_-X^\mu |,
\eeq{Xcom}
where $\mu$ is a target space index.
Note that the density character of $\th$ leads to $\Psi$ and ${\cal F}$
being
densities.

Under $(p+1)$-dimensional diffeomorphisms the components transform as
specified by their density weights, and under Lorentz transformations
according to
their Lorentz charge. The local supersymmetry transformations of the
supergravity fields are 
\ber
\delta e\pp^i &=& \mp i \e^\pm\h\pp^\pm e\pp^i\pm i\e^\mp
(2\h\pp^\mp e\mm^i-\h\mm^\mp e\pp^i)\cr
\delta \h\pp^\pm &=& \d\pp\e^\pm
-\e^\pm\left(\half\d_ie\pp^i\pm
i\h\pp^\mp\h\mm^\mp\pm\omega\pp\right)\cr
&&+\e^\mp\left(2i\h\pp^\mp\h\mm^\pm\mp \ihalf\h\mm^\mp\h\pp^\pm+\ihalf
R\right)\cr
\delta\h\pp^\mp&=&\d\pp\e^\mp\pm {\textstyle{{3i} \over
2}}\e^\pm\h\pp^\mp\h\pp^\pm -\e^\mp\left(\half\d_ie\pp^i\mp
{\textstyle{{3i} \over 2}}\h\pp^\mp\h\mm^\mp\right)\cr
\delta \omega\pp &=&
-\e^\pm\left(\pm i\h\pp^\pm\omega\pp+\h\pp^\mp R\right)\cr
&&\pm\e^\mp\left( 2i\h\pp^\mp\omega\mm - i\h\mm^\mp\omega\pp -
i\rho_\pm\right),\cr
\delta R &=& -\e^+\rho_+ -\e^-\rho_-,
\eer{gtrans}
where $\e$ is the lowest component of the corresponding superfield.
The matter field transformations are
\ber
\delta X^\mu &=& -\e^+\Psi_+^\mu-\e^-\Psi_-^\mu ,\cr
\delta \Psi_\pm^\mu &=& \mp i\e^\pm\d\pp X^\mu\pm\e^\mp{\cal
F}^\mu\mp\ihalf\left(\e^\pm\h\pp^\pm
+\e^\mp\h\mm^\mp\right)\P_\pm^\mu\cr
&&\mp i\e^\pm\h\pp^\mp\Psi_\mp^\mu ,\cr
\delta {\cal F}^\mu &=&
i\e^-\d_=\Psi_+^\mu+i\e^+\d_\+\Psi_-^\mu-\lambda^+\Psi_+^\mu
-\lambda^-\Psi_-^\mu,
\eer{Xtfs}
where  $\lambda ^\pm \is \d_+\d_-\e^\pm|$.

As mentioned in Section 2, one of the consequences of the Bianchi
 identities in conjunction with the constraints \hv{scons} is that
\beq
\d_i\left(e^i_{[\+}e^j_{=]}\right)=0,
\eeq{Afield}
i.e., the $A_i$ field equations of \hv{Eac}.  At the superspace level,
the
locally supersymmetric 
$\s$-model \hv{Spinac} that we consider is therefore a generalization of
\hv{Eac} with the $A_i$ field integrated out. To have unconstrained
fields
in our action we then reintroduce the $A_i$ field as a Lagrange
multiplier in
the component action \hv{Spicom}. The relation
\hv{Afield} is preserved by supersymmetry transformations and
\hv{Spicom} is
thus locally supersymmetric if we take $A_i$ to transform as a singlet. 

Just
like for the purely bosonic case, we may use  
the gauge $\d_ie\pp^i=0$ to choose special coordinates. In
other words, in this gauge the $2D$ Lorentzian ``tangent space'' may be
identified with a subspace of the $(p+1)$-dimensional tangent space to the
bosonic
coordinate space and there are coordinates $\tilde\xi^i \in
\{\tilde\xi^a,\tilde\xi^\+,\tilde\xi^=\}$ with $a=1,\ldots,p-1,$ such that
$e\pp^i={\d\xi^i / \d\tilde\xi{\mbox{\tiny${}^{\stackrel\+ =}$}}}$.
As is customary we also gauge away $\h_A$ using local supersymmetry and 
the typical $2D$-invariance of the action under
\beq
\delta \h_A = \g_A\kappa ,
\eeq{chtr}
where $\kappa (\xi)$ is the transformation parameter.
In this gauge, the field equations that follow from \hv{Spinac} may be
written
\ber
\g\pp\pp &=&\mp i\P_\pm\cdot\d\pp\P_\pm,\cr
\P_\pm\cdot\d\pp X&=&0,\cr
\d_\+\d_=X^\mu &=&0,\cr 
i\d\pp\P_\mp^\mu&=&0 ,
\eer{sfeq}
where we only display the $ {\stackrel \+ =}$ components of the $\delta 
e^i_A$ equations. The equations \hv{sfeq} are precisely those of a
spinning
string in conformal gauge. Again we find a parton picture with the
additional
coordinates labeling a foliation of the world volume. 

\section{Discussion}

We have extended the bosonic results of \cite{lind1} both to include
non-trivial backgrounds for the bosonic theory and to allow for spinning
string
partons. The latter result is necessary if one is to believe in this
limit of $D$-branes which are after all (space-time) supersymmetric.
Ideally,
we should have started from a space-time supersymmetric $D$-brane
action,
derived the strong coupling limit and shown that this could be viewed as
being
built from superstrings. Since space-time and world-sheet
supersymmetric stings are different formulations of the same theory, we
choose instead to construct a world-volume supersymmetrization of our
bosonic
result. We view the emerging picture of spinning string partons as
strong
evidence that the strong coupling limit of $D$-branes can be viewed as a
model
with strings as partons.

In this context it is gratifying to note that it has been shown by Hull
\cite{Hull2} that the only solutions to supergravity at strong coupling
are
strings and particles.

As mentioned in Section 3, our discussion allows for (spinning)
particles as
partons too\footnote{A discussion of $D$-branes as being built from
particles
was recently given in \cite{new}.}. This corresponds to the case when
$e_\+^i$ and
$e_=^i$ are proportional to each other. 

In \cite{lind1} it was shown that for the special case of $p=1$ the
action
\hv{VWac}, (or \hv{Eac}), unifies tensile and tensionless
fundamental strings. The string tension $T$ is an integration constant
for
the $A_i$ equation\footnote{Similar results were previously discussed
in
\cite{blt}.}. This generalizes immediately to the spinning model
presented
here. For $p=1$ \hv{Spicom} leads to the tensile or tensionless
fundamental
spinning string depending on the value of the same integration constant.

We end by some speculative comments on the applicability of our results.
The strong coupling limit of Type IIA strings is described by $11D$
supergravity
(the low energy limit of
$M$-theory). In our strong coupling limit we see nothing that indicates
an
extra dimension appearing. In fact, since we only include the common
(bosonic) sector in the background, we cannot even discriminate between
type IIA and Type IIB, say. We hope to be able to discuss the full
background later using the Hamiltonian description of super
$D$-branes in ref. \cite{BergTown}. 
In any case, for the common sector discussed here the 
background seems to enter in a trivial way in the limit $g_s\to \infty$
for a
single $D$-brane. To see the appearence of an extra dimension, one would
like to be able to take the limit in the $D$-brane action and its background
simultaneously. Apart from this problem, one might perhaps also question
the
description in terms of a standard action in the strong
coupling limit.
Since higher loop quantum effects become more and more important,
perhaps we need to be able to treat 
infinite genus Riemann surfaces for the
open strings describing the $D$-brane fluctuations. 
This might require a novel
description
that includes new effects at such strong coupling.
\bigskip

{\bf \noindent Acknowledgements}
\vskip .2in
\noindent
We are grateful to Ulf Danielsson and Rikard von Unge for
discussions. The work of UL was supported in part by NFR grant
No. F-AA/FU 04038-312 and by NorFA grant No. 96.55.030-O.


\newpage

\begin{thebibliography}{6666}

\newcommand{\np}{{\em Nucl.\ Phys.\ }}
\newcommand{\pr}{{\em Phys.\ Rev.\ }}
\newcommand{\cmp}{{\em Commun.\ Math.\ Phys.\ }}
\newcommand{\pl}{{\em Phys.\ Lett.\ }}

\bibitem{lind1}
U.\ Lindstr\"om and R.\ von Unge, \pl {\bf B403} (1997) 233.

\bibitem{lst1}
U.\ Lindstr\"om,\ B.\ Sundborg and G.\ Theodoridis, \pl
{\bf 258B} (1991) 331.

\bibitem{lr}
 U. Lindstr\"om and M. Ro\v cek, \pl {\bf 271B} (1991) 79.

\bibitem{ul}
U.\ Lindstr\"om, {\em Int. J. Mod. Phys.} {\bf A10} (1988) 2401.

\bibitem{Hull1}
M.~Abou~Zeid and C.~M.~Hull, \pl {\bf 404B} (1997) 264.

\bibitem{Brandt}
F. Brandt, J. Gomis and J. Sim\'on, \pl {\bf B419} (1998) 148.

\bibitem{Lee}
J. Lee, \pr {\bf D57} (1998) 5134.

\bibitem{gulin}
H. Gustafsson and U. Lindstr\"om, ``$2D$-Supergravity in $p+1$
dimensions.''
University of Stockholm preprint USITP-98-12, (1998).

\bibitem{robert}
R. Marnelius, {\em Acta Phys. Pol.} {\bf B13} (1982) 669.

\bibitem{tless}
J.\ Isberg,\ U.\ Lindstr\"om,\ B.\ Sundborg and G.\ Theodoridis, \np {\bf
B411}
(1994) 122.

\bibitem{Kallosh}
R. Kallosh, \pr {\bf D56} (1997) 3515.

\bibitem{Gauntlett}
J. P. Gauntlett, J. Gomis and P. K. Townsend, 
{\em J. High Energy Phys.} {\bf 01} (1998) 003.

\bibitem{jim}
S. J. Gates, Jr. and H. Nishino, {\em Class. Quantum Grav.} {\bf 3}
(1986)
391.

\bibitem{rvnz}
M. Ro\v cek, P. van Nieuwenhuizem and S. C. Zhang, {\em Annals of Phys.}
{\bf 172} (1985) 348.

\bibitem{Hull2}
C. M. Hull, \np {\bf B468} (1996) 113.

\bibitem{new}
H. S. Yang, I. Kim and B.-H. Lee, ``On the Point-like Structure for
Super
p-branes'', hep-th/9806112.

\bibitem{blt}
E.\ Bergshoeff, L.\ A.\ J.\ London and P.\ K.\ Townsend, {\em Class.\
Quantum\
Grav.\ } {\bf 9} (1992) 2545.

\bibitem{BergTown}
E. Bergshoeff and P. K. Townsend, ``Super $D$-branes revisited'',
hep-th/9806112.

\end{thebibliography}
\end{document}